\documentclass[twocolumn,prl,showpacs]{revtex4}

\usepackage{amsmath,amssymb}
\usepackage{graphics}
\usepackage{epsfig}

\begin{document}

\title{Mass Dependence of Ultracold Three-Body Collision Rates} 
\author{J. P. D'Incao and B. D. Esry}
\affiliation{Department of Physics, Kansas State University,
Manhattan, Kansas 66506} 

\begin{abstract}
We show that many aspects of ultracold three-body collisions can be 
controlled by choosing the mass ratio between the collision partners. 
In the ultracold regime, the scattering length dependence of the
three-body rates can be substantially modified from the equal mass
results. We demonstrate that the only non-trivial
mass dependence is due solely to Efimov physics. 
We have determined the mass dependence of the three-body collision
rates for all heteronuclear systems relevant for two-component atomic
gases with resonant $s$-wave interspecies interactions, which
includes only three-body systems with two identical bosons
or two identical fermions. 
\end{abstract} 
\maketitle 

The achievement of quantum degeneracy in ultracold gases with
different atomic species \cite{Exp} has driven experimental studies of
several novel phenomena. The observation of interspecies Feshbach
resonances in boson-fermion mixtures \cite{BFres} allows considerable
flexibility for exploring new regimes by controlling the interspecies 
interactions. Boson-mediated Cooper pairing \cite{bfcooper}, for 
instance, can substantially increase the critical temperature for a
phase transition to the Bardeen-Cooper-Schrieffer (BCS) regime. The
collapse of the fermionic component as well as the phase separation of
both components can be studied \cite{Collapse} along with the
creation of ultracold polar molecules \cite{Baranov}. It is worth
noting that heteronuclear boson-fermion molecules are composite
fermions which allows a new type of crossover between an atomic
Bose-Einstein condensate and a molecular Fermi-type superfluidity
\cite{Kokkelmans}. Heteronuclear boson-boson and fermion-fermion
mixtures have also been studied \cite{Ferrari}, but no Fesh\-bach
resonances have yet been reported.

The magnetic field sensitivity of the hyperfine states is the key 
to controlling the interatomic interactions in ultracold gases.  
By applying an external magnetic field near a diatomic Feshbach
resonance, the $s$-wave scattering length $a$, which characterizes the 
low-energy interatomic interactions, can take any value from the
weakly ($a\rightarrow0$) to the strongly ($|a|\rightarrow\infty$) interacting
limits. Even though two-body loss processes 
can usually be minimized 
by using resonances in the lowest hyperfine states, three-body loss
processes can still be substantial. 
Fortunately, near the resonance, when $|a|\gg r_{0}$ (with $r_{0}$
being the characteristic range of the interatomic interactions)
processes such as vibrational relaxation,
$X\!+\!X^{\ast}_{2}\!\rightarrow\! X\!+\!X_{2}$, three-body
recombination, $X\!+\!X\!+\!X\!\rightarrow \!X\!+\!X^{\ast}_{2}$, and
collision-induced dissociation, $X\!+\!X^{\ast}_{2}\!\rightarrow
\!X\!+\!X\!+\!X$, no 
longer depend on the details of the interactions 
and universal predictions can be made.

Recent experiments have underscored the importance of knowing the $a$
dependence of three-body rates in order to determine the atomic and
molecular lifetimes. In fact, three-body losses have been used to 
locate Feshbach resonances \cite{BFres} and to create ultracold
molecules \cite{Molec}. While general results for threshold
\cite{Threshold} and scattering 
 length scaling laws of three-body equal mass systems
\cite{ScaLenDep} have been obtained, however, there are no similarly general
scaling laws for heteronuclear systems. 
The specific case of recombination in a two-component Fermi gas has been
investigated, though, and found to scale as $a^{6}$
for $a>0$, and minima were predicted as a function of the mass
ratio between the collision partners \cite{PetrovMass}.

In this Letter we demonstrate that the mass ratio has a
large impact on ultracold three-body collisional losses, allowing a
certain degree of control. Using the simple physical picture developed
in Ref.~\cite{ScaLenDep}, extended to include heteronuclear systems,
we have determined that the scattering length scaling laws 
can differ substantially from the equal mass results \cite{ScaLenDep}. 
For instance, in a system with two identical fermions 
that are much heavier than the third atom, relaxation of weakly bound  
heteronuclear molecules scales approximately as $a^{-7}$ ---
an even stronger suppression than the $a^{-3.33}$ scaling found when
all three atoms have equal mass \cite{ScaLenDep,Petrov}. 
This scaling was derived in Ref.~\cite{Petrov}
to explain the long molecular lifetimes observed experimentally for
molecules formed of fermions in different spin states. It was their
long lifetimes that made further experiments with these molecules
feasible \cite{Molec,LongLivedMol}. The stronger suppression found
here might open other experimental avenues. 

In our picture, the mass dependence enters via the three-body
effective potentials and is associated with Efimov physics
\cite{Efimov}. In fact, we demonstrate here that the only non-trivial
mass dependence is due solely to Efimov physics. 
We determine the mass dependence of the collision rates for
three-body systems relevant to all two-component atomic gases with
resonant interspecies $s$-wave interactions, which includes systems
with two identical bosons or two identical fermions. 

For short-range two-body interactions, the ultracold behavior of
three-body systems can be derived from three-body
effective potentials and couplings \cite{ScaLenDep} which,
in the adiabatic hyperspherical representation, are determined from the
adiabatic equation \cite{Esry02}, 
\begin{equation}
H_{\rm ad}(R,\Omega)\Phi_\nu(R;\Omega)=U_\nu(R) \Phi_\nu(R;\Omega).
\label{adeq}
\end{equation}
\noindent
This equation is obtained from the full Schr\"{o}dinger equation by
fixing the hyperradius $R$, leaving only dynamics in the hyperangles
$\Omega$ through $H_{\rm ad}$. By expanding the total wave function on
the adiabatic basis $\Phi_\nu$, the Schr{\"o}dinger equation (in atomic
units) is reduced to: 
\begin{eqnarray} 
\left[-\frac{1}{2\mu}\frac{d^2}{dR^2}+W_{\nu}\right]F_{\nu}
+\sum_{\nu'\neq\nu} V_{\nu\nu'} F_{\nu'}=E F_\nu, 
\label{radeq}
\end{eqnarray}
\noindent
where $E$ is the total energy,
$F_{\nu}$ is the hyperradial wave function, and $\nu$ is a collective
index that represents all quantum numbers necessary to label each
channel. 
In the present case, the three-body reduced mass is 
$\mu$=$m/\sqrt{\delta(\delta+2)}$ where $m$ is the mass of the
distinguishable particle and $\delta$ is the ratio of the 
different atom mass to the
identical atom mass.
This equation describes the collective radial motion under 
the influence of the effective potential $W_{\nu}$ as well as any inelastic 
transitions controlled by the nonadiabatic couplings $V_{\nu\nu'}$. 

In the limit $|a| \gg r_{0}$ the effective potentials $W_{\nu}$ depend
crucially on Efimov physics \cite{Efimov,ScaLenDep}. In
Ref.~\cite{ScaLenDep}, we described a scheme that uses this fact to
classify all equal-mass three-body systems with 
resonant $s$-wave pairwise interactions. In this scheme, each system
falls into one of two categories: those with an attractive dipole ($R^{-2}$)
potential in the range $r_{0} \ll R \ll |a|$ and those only with
repulsive dipole potentials.
Which category a particular system falls into depends on its symmetry
$J^{\pi}$ (total angular momentum $J$ and parity $\pi$) and
the identical particle permutational symmetry, i.e., whether they are
bosons or fermions. 
For heteronuclear systems, this basic classification scheme still
holds, and the extension of the analysis of the
three-body rates requires only
slight modifications which, in turn, come almost 
entirely from the $\delta$ dependence of the dipole potential strengths.

For each category, there are three distinct regions in 
$R$ that characterize the effective potentials (for a schematic
picture, see Fig.~1 in Ref.~\cite{ScaLenDep}): $R\lesssim r_0$, 
$r_0$$\ll$$R$$\ll$$|a_\delta|$, and $R$$\gg$$|a_{\delta}|$
where $a_\delta$=$[\sqrt{\delta(\delta+2)}/(\delta+1)]^{1/2}a$.   
Replacing $a$ by $a_\delta$ in these definitions is the first
mass-dependent modification of our previous analysis.
In the asymptotic region, $R$$\gg$$|a_{\delta}|$,
the potentials can be derived analytically~\cite{NielsenReview}. They
are associated with molecular channels, which represent atom-molecule
scattering, and with three-body continuum channels, which represent
collisions of three free atoms, and are given respectively by   
\begin{equation} 
W_{\nu}={E}_{vl'}\!+\!\frac{l(l\!+\!1)}{2\mu R^2}
~~\mbox{and}~~  W_{\nu}=\frac{\lambda(\lambda\!+\!4)+\!15/4}{2\mu R^2}.
\label{bcch}
\end{equation} 
The molecular bound state energy ${E}_{vl'}$ is labeled by the
ro-vibrational quantum numbers $v$ and $l'$; $l$ is the atom-molecule
relative angular momentum; and $\lambda$ 
labels the eigenstates of the hyperangular kinetic energy. 

For $r_{0}$$\ll$$R$$\ll$$|a_{\delta}|$, the potentials for the molecular and
continuum channels [Eq.~(\ref{bcch})] are modified due to Efimov
physics, establishing our classification scheme. In the first category,
an attractive dipole potential occurs in the highest vibrationally
excited $s$-wave molecular channel for $a > 0$ and in the lowest
continuum channel for $a < 0$. The potentials for all higher-lying
channels are repulsive. These potentials are conveniently parametrized
by the coefficients $s_0$ and $s_{\nu}$:  
\begin{eqnarray}
W_{\nu}(R)=-\frac{s^{2}_{0}+\frac{1}{4}}{2\mu R^2}
\hspace{0.25cm}\mbox{and}\hspace{0.25cm}
W_{\nu}(R)=\frac{s^{2}_{\nu}-\frac{1}{4}}{2\mu R^2}.
\label{bcchefbosonic} 
\end{eqnarray}
In the second category, the potentials for the weakly bound molecular
channel and the continuum channels are repulsive and
parametrized by coefficients $p_0$ and $p_{\nu}$:  
\begin{eqnarray}
W_{\nu}(R)=\frac{p^{2}_{0}-\frac{1}{4}}{2\mu R^2}
\hspace{0.25cm}\mbox{and}\hspace{0.25cm}
W_{\nu}(R)=\frac{p^{2}_{\nu}-\frac{1}{4}}{2\mu R^2}.
\label{bccheffermionic} 
\end{eqnarray}
In both categories, however, deeply bound molecular channels are
essentially independent of $a$. In the above equations, the
coefficients $s_0$, $s_{\nu}$, $p_0$, and $p_{\nu}$ 
depend on the number of resonant pairs, the number of identical
particles, and the mass ratio $\delta$ between collision partners 
\cite{Efimov}. 
For $R \lesssim r_{0}$, the potentials depend on the details
of the interatomic interactions and can lead to resonance effects due
to three-body shape or Feshbach resonances.

In the adiabatic hyperspherical representation, inelastic
transitions at ultracold temperatures proceed via tunneling in the
initial collision channel to where the coupling $V_{\nu\nu'}$ peaks
--- the coupling, of course, drives the transition. 
We have previously shown that a simple WKB approximation to the 
tunneling probability through the potentials given by
Eqs.~(\ref{bcch})-(\ref{bccheffermionic}) is sufficient to determine
both the threshold and the scattering length scaling laws for
equal mass three-body collisions \cite{ScaLenDep}. 
For heteronuclear systems, $a_{\delta}$, $s_0$, $s_{\nu}$, $p_0$, and
$p_{\nu}$ generate the only non-trivial mass dependence in the
three-body rates and follow directly from Efimov's analysis.

For systems and symmetries from our first category [with
an attractive potential (\ref{bcchefbosonic})],
relaxation for $a$$>$$0$ is
\begin{align}
V_{\rm rel}\!\propto 
\!{\mu}^{l-1}\!E_{\rm coll}^{l}
\!
\frac{\sinh(2\eta)}{\sin^2\!\left[s_{0}\ln({a_{\delta}}/{r_{0}}) 
\!+\!\Phi\right]\!+\!\sinh^2(\eta)}
a_{\delta}^{2l+1};
\label{vibrelApos}
\end{align}
\noindent
and for $a$$<$$0$, $V_{\rm rel}$$\propto$$
{\mu}^{l-1}E_{\rm coll}^{l}{r_{0}^{2l+1}}$.
$E_{\rm coll}$=$E\!-\!E_{vl'}$ is the collision energy,
$\Phi$ is an unknown short-range phase~\cite{ScaLenDep}, and $\eta$ labels parameters
related to the inelastic transition probability at small distances
\cite{BraatenReview}. 
These parameters 
can depend on the masses nontrivially and can lead 
to resonance effects, depending upon details of the interactions.  
Recombination for $a$$>$$0$ and $a$$<$$0$ is given by
\begin{align}
K_{3}&\!\propto \!
{\mu}^{\lambda-1}{E}^{\lambda}\!
\left[\sin^{2}[s_{0}\ln(\frac{a_\delta}{r_0})\!+\!\Phi]\!+\!
A_{\eta}(\frac{r_0}{a_\delta})^{2s_{\nu}}\right]\!
{a_{\delta}^{2\lambda+4}}, \nonumber\\
K_{3}&\!\propto \!
{\mu}^{\lambda-1}E^{\lambda}
\frac{\sinh(2\eta)}{\sin^2\left[s_0\ln({|a_{\delta}|}/{r_{0}})
\!+\!\Phi\right]\!+\!\sinh^2(\eta)}
|a_{\delta}|^{2\lambda+4},
\label{RecombAneg}
\end{align}
\noindent
For systems and symmetries from our second category [with
a repulsive potential (\ref{bccheffermionic})], relaxation
for $a>0$ is
\begin{eqnarray} 
V_{\rm rel}\propto 
{\mu}^{l-1}E_{\rm coll}^l
({r_{0}}/{a_{\delta}})^{2p_{0}}{a_{\delta}^{2l+1}},
\label{vibrelBpos}
\end{eqnarray}
\noindent 
while it is $V_{\rm rel}\propto {\mu}^{l-1}E_{\rm coll}^{l} 
{r_{0}^{2l+1}}$ for $a<0$. Recombination for $a>0$ and
$a<0$ is given by 
\begin{align}
K_{3}\!&\propto\!
{\mu}^{\lambda-1}E^{\lambda}\!
\left[1\!+\!A_{\eta}(\frac{r_{0}}{a_{\delta}})^{2p_{0}}
\!+\!B_{\eta}(\frac{r_{0}}{a_{\delta}})^{2p_{\nu}}\right]\!
a_{\delta}^{2\lambda+4}, \nonumber\\
K_{3}&\propto 
{\mu}^{\lambda-1}E^{\lambda}
({r_{0}}/{|a_{\delta}|})^{2p_{0}}
|a_{\delta}|^{2\lambda+4}. 
\label{RecombBneg}
\end{align}

\begin{figure}[htbp]
\includegraphics[width=1.7in,height=3.25in,angle=270,clip=true]{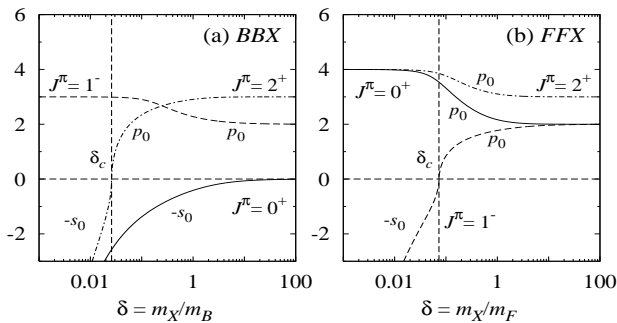} 
\caption{
  Mass dependence of $s_{0}$ and $p_{0}$, for (a) two identical
  boson and (b) two identical fermion systems. \label{Fig1} 
  }  
\end{figure}

The mass dependence of $s_0$ and $p_0$ can substantially modify the
$a$ dependence of the three-body rates. 
The coefficients $s_{\nu}$ and $p_{\nu}$, however,
do not affect the scaling laws.
Figure~\ref{Fig1} shows $s_{0}$ and $p_{0}$
for systems with two identical bosons ($BBX$, $\delta$=$m_{X}/m_{B}$)
and with two identical fermions ($FFX$, $\delta$=$m_{X}/m_{F}$).
These coefficients are determined analytically following
Ref.~\cite{Efimov} after proper symmetrization. 
It is important to notice that for $J>0$ the effective 
potentials can change from attractive to repulsive and vice-versa,
changing the category in which the system is classified, and thus 
modifying its collisional properties. 
It happens, for instance, for $2^+$ $BBX$ systems at
$\delta_{c}\approx0.0259$ [Fig.~\ref{Fig1}(a)] and
for $1^-$ $FFX$ systems at $\delta_{c}\approx0.0735$
[Fig.~\ref{Fig1}(b)].  

For bosonic systems $BBX$, the dominant contribution for relaxation
and recombination is $J^{\pi}=0^+$ with $l=0$ and
$\lambda = 0$, and is described by
Eqs.~(\ref{vibrelApos})-(\ref{RecombAneg}) due to the presence of an
attractive dipole potential.
In this case, the variations of $s_{0}$ with 
$\delta$ [Fig.~\ref{Fig1}(a)] changes the number of Efimov states. These
variations manifest themselves in the three-body rates through the
locations of the minima and peaks in
Eqs.~(\ref{vibrelApos})-(\ref{RecombAneg}). We note that the mass
dependence in this case does not modify the scaling laws.  

In contrast, for fermionic systems $FFX$, the power law behavior does
change with mass for the dominant contributions to relaxation
($J^{\pi}=0^+$ with $l=0$) and recombination ($J^{\pi}=1^-$ with
$\lambda=1$). These changes follow from the mass dependence of the
repulsive dipole potential due to $p_{0}$ [Fig.~\ref{Fig1}(b)]
combined with Eqs.~(\ref{vibrelBpos})-(\ref{RecombBneg}).   
It follows that relaxation for $FX+F$ collisions scales as
$a^{1-2p_{0}}$ for $a > 0$ and is suppressed for all $\delta$ since 
2$\le$$p_{0}$$\le$4 [Fig.~\ref{Fig1}(b)], approaching its greatest
suppression, $a^{-7}$,
as $\delta\rightarrow 0$ and its least, $a^{-3}$, as
$\delta\rightarrow\infty$. Suppression of $FX+F$ collisions can thus 
be much stronger than $a^{-3.33}$, found for $\delta=1$ $FF'+F$ 
collisions of fermionic atoms in different spin states
\cite{ScaLenDep,Petrov}. 
Extremely long-lived heteronuclear molecules in ultracold
boson-fermion and fermion-fermion mixtures might thus be possible. 
The modifications to the recombination scaling law occurs 
for $a<0$ and $\delta\ge0.0735$ where it scales as $a^{6-2p_{0}}$
with $0\le p_{0}\le 2$ [Fig.~\ref{Fig1}(b)], otherwise it scales as 
$a^6$, leading to an asymmetry between $a>0$ and $a<0$. The strongest
asymmetry is obtained when $\delta\rightarrow\infty$, where
recombination scales as $a^6$ and $|a|^2$, respectively, so that greater
collisional stability against recombination might be expected for
$a<0$.  

As mentioned above, recombination for $FFX$ systems with
$\delta \ge 0.0735$ was studied in Ref.~\cite{PetrovMass} and found to
scale as $a^6$ for $a > 0$ which agrees with our prediction.
It was also predicted that the recombination rate oscillates as a
function of $\delta$ with zeros at $\delta = 0.0735$, $0.1160$, and as
$\delta \rightarrow \infty$ (note that the mass ratio in
\cite{PetrovMass} is $1/\delta$). 
These zeros were ascribed to a decoupling of
the three-body continuum and bound channels \cite{PetrovMass}, leading
to elastic scattering only, and have also been interpreted as interference
effects \cite{BraatenReview}. 
In our formulation, however, no such zeros are predicted. In fact, in
this range of $\delta$, the potentials for both continuum and bound
channels, Eq.~(\ref{bccheffermionic}), are repulsive and interference
effects are expected to be suppressed by tunneling.  
Further, we have noticed through representative numerical calculations
that the continuum and bound channels never decouple and that the
nonadiabatic coupling peaks at $R \approx r_{0}$ and $R \approx a$, in
accord with our model \cite{ScaLenDep}. While we have observed minima
in our calculations, they do not correspond to the minima discussed in
Ref.~\cite{PetrovMass}. They are most likely a consequence of using a
two-body potential with finite $a$, and we expect them to disappear in
the limit $a\rightarrow\infty$ considered in
\cite{PetrovMass}. Preliminary numerical results support this
expectation.

Table \ref{TabRates} summarizes the threshold and scattering length 
scaling laws for the three-body rates, including dissociation $D_{3}$. 
The three dominant partial waves (as determined by their energy
dependence) are shown for each process. 
\begin{table}[htbp]
\begin{ruledtabular}
\begin{tabular}{ccccccc}
 & & \multicolumn{2}{c}{$V_{\rm rel}$} &  \multicolumn{3}{c}{$K_{3}$ $(D_{3})$} \\ 
 & $J^{\pi}$ & $E$  & $a>0$ & $E$ & $a>0$ & $a<0$   \\ \hline
$BBX$ &$0^+$ & {\bf const} &\boldmath{$a$} & {\bf const}\boldmath{$(E^{2})$} &\boldmath{$a^{4}$} & \boldmath{$|a|^{4}$}   \\
    &$1^-$ & $E_{\rm coll}$ & $a^{3-2p_{0}}$ & $E$$(E^{3})$ & $a^{6}$  & $|a|^{6-2p_{0}}$ \\
    &$2^+$ & $E_{\rm coll}^{2}$ & $a^{5}$,$a^{5-2p_{0}}$ & $E^{2}$$(E^{4})$ & $a^{8}$,$a^{8}$ & $|a|^{8}$,$|a|^{8-2p_{0}}$\\
$FFX$ &$0^+$ & {\bf const} & \boldmath{$a^{1-2p_{0}}$} 
    & $E^{2}$$(E^{4})$ & $a^{8}$  & $|a|^{8-2p_0}$ \\
    &$1^-$ & $E_{\rm coll}$ & $a^{3}$,$a^{3-2p_{0}}$&
 \boldmath{$E$}\boldmath{$(E^{3})$} & \boldmath{$a^{6}$,$a^{6}$} &
 \boldmath{$|a|^{6}$},\boldmath{$|a|^{6-2p_{0}}$}\\
    &$2^+$ & $E_{\rm coll}^{2}$ & $a^{5-2p_{0}}$ & $E^{2}$$(E^{4})$ & $a^{8}$  & $|a|^{8-2p_0}$ \\
\end{tabular}
\end{ruledtabular}
\caption{Threshold and scattering length scaling laws for three-body
  rates. For $2^+$ $BBX$ and $1^-$ $FFX$ systems the rates are given for
  $\delta<\delta_{c}$ and $\delta>\delta_{c}$. 
  Boldface indicates the leading contribution at threshold.  
  \label{TabRates}} 
\end{table}
The table thus indicates the
terms expected to be important for finite energies or
$|a|\rightarrow\infty$. For $BBX$ systems, these contributions are not
expected to be important for relaxation with $a>0$ or recombination
with $a<0$. For $FFX$ systems, however, the higher partial waves are
comparatively more important.

In two-component ultracold atomic gases with resonant interspecies
interactions, the important three-body processes are those that
involve both atomic species, assuming that intraspecies interactions 
are not resonant.
In this case, there are only two relevant 
systems and the competition between the collision rates for each
system as well as the density of each species determines the 
atomic and molecular lifetimes. 
In boson-fermion mixtures, for instance, relaxation for $BF+F$
collisions decreases with $a$ while for $BF+B$ it increases.
In this case, if all bosons are 
bound in $BF$ molecules, the molecules are
expected to be long-lived. Otherwise, the molecules will be rapidly 
quenched by collisions with the bosonic atoms. 

Vibrational relaxation, as we have shown, can be controlled by
choosing the atomic species according to their mass and consequently
the molecular lifetime can be substantially modified.
We show in Table~\ref{TabRatesMixtures} the suppression predicted
for $BF+F$ collisions of commonly used alkali
atoms. The table includes systems used in recent
experiments: $^{23}$Na-$^{6}$Li and $^{87}$Rb-$^{40}$K
\cite{BFres}, as well as $^{7}$Li-$^{6}$Li \cite{Kokkelmans}. We have
included H due to the prospects for using it for sympathetic
cooling~\cite{Cote}. 
Molecule-molecule collisions also contribute to the molecular lifetime
\cite{Petrov}, but in boson-fermion mixtures the molecules are
composite fermions so that only $p$-wave molecule-molecule collisions
occur. In this case, molecule-molecule collisions are suppressed for
ultracold temperatures, and atom-molecule collisions are expected to be
dominant. 
\begin{table}[htbp]
\begin{ruledtabular}
\begin{tabular}{lclclc}
\multicolumn{1}{c}{$B$-$F$} & \multicolumn{1}{c}{$V_{\rm rel}^{BF+F}$} &
\multicolumn{1}{c}{$B$-$F$} & \multicolumn{1}{c}{$V_{\rm rel}^{BF+F}$} &
\multicolumn{1}{c}{$B$-$F$} & \multicolumn{1}{c}{$V_{\rm rel}^{BF+F}$}
\\ [0.05in]\hline
{$^{133}$Cs-$^{6}$Li} & $a^{-3.00}$ &
{$^{23}$Na-$^{6}$Li}  & $a^{-3.05}$ & 
{$^{}$H-$^{6}$Li}     & $a^{-4.89}$ \\
{$^{133}$Cs-$^{40}$K} & $a^{-3.06}$ &
{$^{23}$Na-$^{40}$K}  & $a^{-3.63}$ & 
{$^{}$H-$^{40}$K}     & $a^{-6.85}$ \\
{$^{87}$Rb-$^{6}$Li}  & $a^{-3.01}$ &
{$^{7}$Li-$^{6}$Li}   & $a^{-3.27}$ &
                      &             \\
{$^{87}$Rb-$^{40}$K}  & $a^{-3.12}$ &
{$^{7}$Li-$^{40}$K}   & $a^{-4.83}$ &
                      &  
\end{tabular}
\end{ruledtabular}
\caption{Scattering length scaling laws for relaxation ($a>0$) of
  $BF+F$ collisions in boson-fermion
  mixtures based solely on mass. \label{TabRatesMixtures}}    
\end{table}

In this Letter, we have explored the mass dependence of the ultracold
three-body collision rates for systems with two identical particles that
interact resonantly with a third particle. 
We have derived the energy and scattering length dependence for all
three-body inelastic collision rates that apply to two-component
ultracold quantum gases with resonant interspecies interactions.  
In the process, we have demonstrated that the mass dependence in
three-body collisions is intimately related with Efimov physics.  
In bosonic systems, the mass dependence affects the features
due to Efimov physics but leaves the dominant scattering length
scaling laws unchanged. In fermionic systems, however, the mass
dependence modifies the scattering length scaling laws substantially
from the equal mass results.  
The stronger suppression found for relaxation of weakly bound
heteronuclear molecules in boson-fermion mixtures, suggests the
possibility of long-lived heteronuclear fermionic
molecules.     
 
This work was supported by the National Science Foundation.

\end{document}